# Positive Personas: Integrating Well-being Determinants into Personas

Irawan Nurhas [1], Stefan Geisler [1], Jan Pawlowski [1]

Institute of Positive Computing, Ruhr West University of Applied Science[1]

**Abstract**

System design for well-being needs an appropriate tool to help designers to determine relevant requirements that can help human well-being to flourish. Personas come as a simple yet powerful tool in the early development stage of the user interface design. Considering well-being determinants in the early design process provide benefits for both the user and the development team. Therefore, in this short paper, we performed a literature study to provide a conceptual model of well-being in personas and propose positive design interventions in personas' creation process.

## 1 Introduction and related work

The upcoming trend towards positive design to promote human well-being in the development of an information system (Pawlowski et al., 2015) brings an essential movement to propose an appropriate hands-on method for a designer and integrating well-being goals in the early stage of the design process (Desmet & Pohlmeyer, 2013). Through literature study, we summarized that well-being is any condition that causes positive emotional reactions in the short- or long-term for an individual reason. Consequently, it is important to identify artifacts start from the beginning that can help flourishing user well-being. Parallel to the positive computing movement, Cooper et al. (2014) recommends the use of personas as starting point to model the usage behavior in the process of user interface design. Despite the importance of well-being and personas in the initial development process. To our knowledge, there is still a lack of integration of well-being into personas that can be used as the bridge between research and design, and as the communication tool between cross-functional departments in the development phase of system design. Therefore, this study aims to answer the research question about how a well-being goal-oriented design can be integrated into the personas tool. To answer this questions, we have performed a literature study and provided a well-being model for personas.

For the aim of the study, the Human-Artifact Model (HAM) and Positive Activity Model (PAM) were selected to build positive personas. HAM was chosen because it can be used to describe the relation in technology-usage activities between a usage artifact and the human aspect (Bødker & Klokmose, 2011), whether PAM can be used to improve human well-being by taking attention to the activity and personal features (Lyubomirsky & Layous, 2013). Moreover, both models facilitate fundamental questions and share similarity (why, what, and how?) related to user activity.

## 2 Methodology

In this study, we utilized design science research (Peffers et al., 2007) by focusing on the structured literature review (Webster & Watson, 2002) to explore and build an initial model for a new topic in personas. We filtered inappropriate articles and analyzed 89 articles to build our model. As a starting point, we have extracted the modeling process of well-being by Konu & Rimpela (2002), and we followed the process as our guideline to construct the model. The summarized processes are 1) identification of well-being related activity in the personas; 2) selection of HAM and PAM as the basic model; 3) analyzing the similarities of the model and finding missing critical aspects of the context, based on the selected model.

## 3 Conceptualizing positive personas

The selection of well-being determinant of each artifact in the positive personas (see Figure 1) were based on the literature finding by answering combined questions between HAM and PAM or combination between PAM's questions (when and who) with an object in personas creation process (Cooper et.al, 2014). Figure 1 shows our proposed model for integrating well-being determinant into personas attributes.

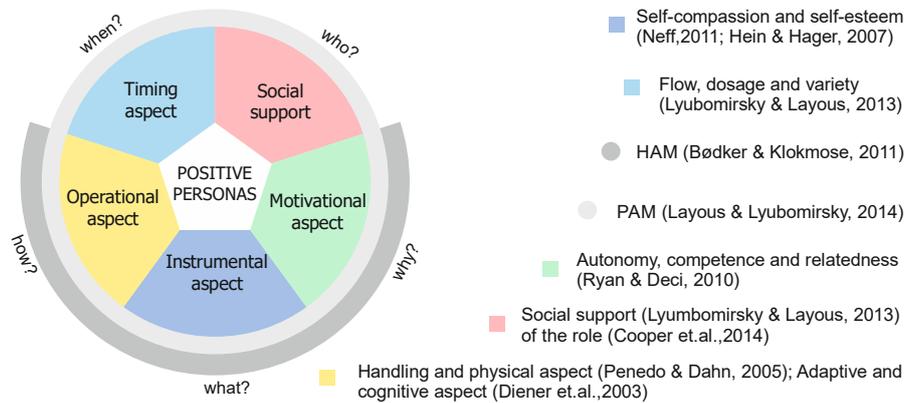

*Figure 1: Positive personas*

Next, we adopted the creation process of personas by Cooper et. al. (2014) and related the object of each step in the process with our positive personas aspect. (e.g., role ↔ social support, behavior ↔ motivation, pattern ↔ timing, the user characteristic & user goals ↔ operational aspect & instrumental aspect). The complete proposed intervention can be seen in Table 1.

| Personas creation process (cooper et al, 2014, p 82) | Proposed intervention based on positive personas |
|---|---|
| Group interview subject by role | Identify the social support of each role |
| Identify behavioral variables and significant pattern of behavior | Classify behaviors that can improve the feeling of competence, autonomy, and relatedness. Identify frequency, sequence, and variety of usage behavior. |
| Synthesize characteristics and define goals | Identify the past and current physical or cognitive conditions that can support adoption of the system and can facilitate a high level of self-compassion as well as self-esteem in obtaining the usage goals. |

*Table 1: adoption of positive personas*

Lastly, we tried to apply our model to build a short sample of positive personas. We framed our personas by extracting information based on published literature (Loe, 2010) and "Ruth" was chosen as our primary personas. The comparison of personas can be seen in Figure 2.

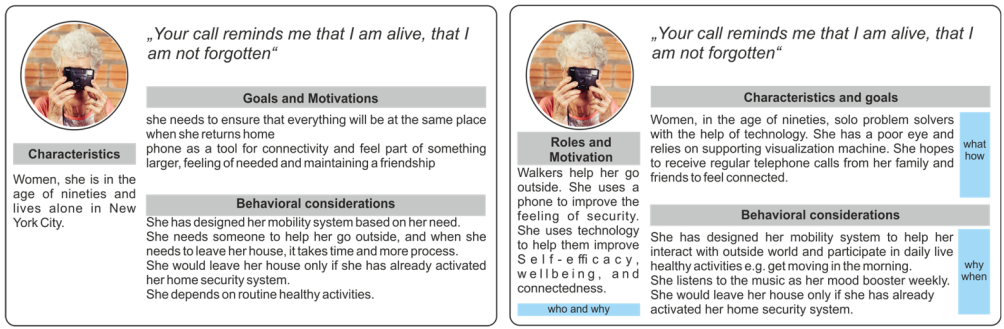

*Figure 2: comparison of personas. Without positive personas (left) and with positive personas (right)*

# 4    Discussion and conclusion

Our positive personas extend the current focus of personas to well-being mediators that drive the usage goals and behaviors. By utilizing positive personas, the designers need to relate the personas attributes to well-being determinants that can facilitate collaboration among the team and to communicate design decision as well evaluate selected requirements from well-being based views. Further testing is necessary to improve the validity of our proposed positive personas, which will be fully implemented and tested on further case studies and projects. Once such models are developed, they could be re-used to help the designer limit the field of interpretation and collaborate with the cross-functional department on the group working.

# References

Bødker, S., & Klokmose, C. N. (2011). The human–artifact model: An activity theoretical approach to artifact ecologies. Human–Computer Interaction, 26(4), 315-371.

Calvo, R. A., & Peters, D. (2014). Positive computing: technology for wellbeing and human potential. MIT Press.

Cooper, A., Reimann, R., Cronin, D., & Noessel, C. (2014). About face: the essentials of interaction design. John Wiley & Sons.

Desmet, P. M., & Pohlmeyer, A. E. (2013). Positive design: An introduction to design for subjective well-being. International Journal of Design, 7(3).

Diener, E., Oishi, S., & Lucas, R. E. (2003). Personality, culture, and subjective well-being: Emotional and cognitive evaluations of life. Annual review of psychology, 54(1), 403-425.

Hein, V., & Hagger, M. S. (2007). Global self-esteem, goal achievement orientations, and self-determined behavioural regulations in a physical education setting. Journal of sports sciences, 25(2), 149-159.

Konu, A., & Rimpelä, M. (2002). Well-being in schools: a conceptual model. Health promotion international, 17(1), 79-87.

Layous, K., & Lyubomirsky, S. (2014). The how, why, what, when, and who of happiness: Mechanisms underlying the success of positive activity interventions. Positive emotion: Integrating the light sides and dark sides, 473-495.

Loe, M. (2010). Doing it my way: old women, technology and wellbeing. Sociology of health & illness, 32(2), 319-334.

Lyubomirsky, S., & Layous, K. (2013). How do simple positive activities increase well-being? Current Directions in Psychological Science, 22(1), 57-62.

Neff, K. D. (2011). Self-compassion, self-esteem, and well-being. Social and personality psychology compass, 5(1), 1-12.

Pawlowski, J. M., Eimler, S. C., Jansen, M., Stoffregen, J., Geisler, S., Koch, O. & Handmann, U. (2015). Positive Computing. Business & Information Systems Engineering, 57(6), 405-408.

Peffers, K., Tuunanen, T., Rothenberger, M. A., & Chatterjee, S. (2007). A design science research methodology for information systems research. Journal of management information systems, 24(3), 45-77.

Penedo, F. J., & Dahn, J. R. (2005). Exercise and well-being: a review of mental and physical health benefits associated with physical activity. Current opinion in psychiatry, 18(2), 189-193.

Ryan, R. M., & Deci, E. L. (2000). Self-determination theory and the facilitation of intrinsic motivation, social development, and well-being. American psychologist, 55(1), 68.

Webster, J., & Watson, R. T. (2002). Analyzing the past to prepare for the future: Writing a literature review. MIS quarterly, xiii-xxiii.